\documentclass[aps,prl,twocolumn,showpacs,groupedaddress]{revtex4}
\usepackage{graphicx}

\begin{document}

%\draft
\title{Noise-assisted preparation of entangled atoms}
\author{X. X. Yi, C. S. Yu,  L. Zhou, and H. S. Song}
\affiliation{Department of physics, Dalian University of Technology, Dalian 116024, China}

\date{\today}

\begin{abstract}
We discuss the generation of entangled states of two two-level
atoms inside an optical cavity. The cavity mode is supposed to be
coupled to a white noise with adjustable intensity. We describe
how the entanglement between the atoms inside the cavity arise in
such a situation. The entanglement is maximized for intermediate
values of the noise intensity, while it is a monotonic function of
the spontaneous rate. This resembles the phenomenon of stochastic
resonance and sheds more light on the idea to exploit white noise
in quantum information processing.
\end{abstract}

\pacs{ 03.67.-a, 03.67.-Hz} \maketitle

Entanglement shared between distant sites is a valuable resource
for quantum information processing \cite{bennett, nielsen,
plenio1}. Stimulated by this discovery, there are a lot of
attention have been focused on how to create, manipulate and
exploit entanglement from both sides of experimental and
theoretical studies \cite{raimond}. From a practical point of
view, the main task to create and exploit entanglement is how to
minimize the impact of noise that is not able to be isolated from
our system in any real experimental scenario. This interaction
between the system of interest and the noise which usually models
surroundings of the system results in a decoherence process. As a
consequence the entangled system may end up in a mixed state that
would be no longer useful for any quantum information processing.
It is therefore important for practical realization of quantum
information processing protocols to engineer mechanisms to
prevent, minimize, or use the impact of environmental noise.

Numerous proposals have been made for preventing, minimizing or
using the environmental noise, for example, loop control
strategies, that use an ancillary system coupling to  the quantum
processor to better the performance of the
proposals\cite{wiseman,mancini}, quantum error correction
\cite{shor} uses redundant coding to protect quantum states
against noisy environments. This procedure is successful as long
as the error rate is sufficiently small. It wastes a number of
qubits and quantum gates, and then limit its implementation by
present available technology. A more economic approach consists of
exploiting the existence of so-called decoherence-free subspace
that are completely insensitive to specific types of noise
\cite{palma}. This approach tends to require fewer additional
resources, but is only applicable in specific situations. The
seminal idea that dissipation can assist the generation of
entanglement has been put forward recently \cite{plenio2,
bose,plenio3}. In a system consisting of two distinct leaky
optical cavities, it was shown that the entanglement is maximized
for intermediate values of the cavity damping rates and the
intensity of the white noise, vanishing both for small and for
large values of these parameters \cite{plenio3}. In fact, this
idea appeared first in Ref.\cite{plenio2} for two atoms inside an
optical cavity and it shows that cavity decay can assist the
preparation of maximally entangled atoms, without cavity decay,
the reduced state of the two-atom system would be in an
inseparable mixture at all times, but not in a maximally entangled
one.

In the latter case, the photon leakage leads to the undesired
parts of the global wave-function to decay and therefore such
terms are eliminated for sufficiently large times. This requires
that the overlapping between the initial state and the desired
maximally entangled state must not be zero, hence one of the two
atoms prepared  in its excited state initially is necessary. This
point makes our proposed scheme here for preparation of entangled
two atoms different from the above one.

In this letter, we put the idea in Ref. \cite{plenio3} forward to
a two-atom system, we use noise to play a constructive role in
quantum information processing. We will concentrate on the problem
of creating entangled atoms when only incoherent sources are
available and demonstrate that, indeed, controllable entanglement
can arise in this situation.

Our system consists of two two-level atoms inside a leaky optical
cavity. As depicted in figure 1,
\begin{figure}
\includegraphics*[width=0.95\columnwidth,
height=0.6\columnwidth]{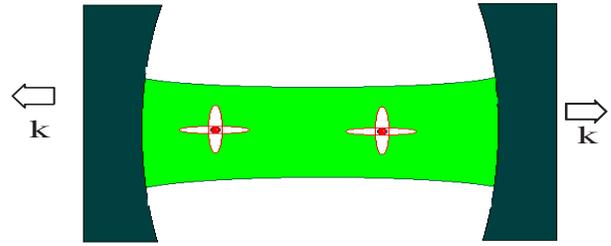} \caption{Schematic diagram for
the preparation of entangled two atoms. The atom-cavity system is
prepared initially in the ground state $|g\rangle_a|g\rangle_b$
with no photon populated in the cavity. The atoms become entangled
and the entanglement is maximized for intermediate values of the
noise intensity. } \label{fig1}
\end{figure}
we will refer to atom $a$  and atom $b$ when the context requires
us to differentiate them, but otherwise they are supposed to be
identical. We denote the atomic ground and excited states by
$|g\rangle_i$ and $|e\rangle_i$, respectively,  and call $2\Gamma$
($\Gamma_a=\Gamma_b=\Gamma$) the spontaneous emission rate from
the upper level. We assume that the distance between the atoms is
much larger than an optical wavelength,  therefore dipole-dipole
interaction can be neglected. The cavity mode is assumed to be
resonant with the atomic transition frequency, and we will denote
the cavity decay rate by $2\kappa$. For the sake of generality we
allow the coupling between each atom and the cavity mode, $g_i$,
to be different. We suppose the cavity is driven by an external
thermal field(white noise) whose intensity will be characterized
in terms of an effective photon number $n_T$, but we do not want
to specify the noise. The relaxation of the atom-cavity system can
take place through two channels, at rate $2\kappa$ (cavity decay)
and $2\Gamma$ (spontaneous decay). The master equation governing
the time evolution of the global system is given by (setting
$\hbar=1$)
\begin{equation}
\dot{\rho}=-i[H,\rho]+{\cal L} (\rho),
\end{equation}
where the Hamiltonian $H$ describes the internal energies of the
cavity and the two atoms as well as the atom-cavity coupling. The
Liouvillean ${\cal L }(\rho)$ describes the atom decay and the
interaction of the cavity mode with the noise. As no external
coherent driving is present, the Hamiltonian reads
\begin{equation}
H=\frac{\omega}{2}\sigma^z_a+\frac{\omega}{2}\sigma_b^z+\omega_fa^{\dagger}a+\sum_{i=a,b}
g_i(|g\rangle_i\langle e|a^{\dagger}+h.c.),
\end{equation}
where $a$ represent the cavity mode with frequency $\omega_f$. The
Liouvillean is given by
\begin{eqnarray}
{\cal L}(\rho)&=&-\kappa(n_T+1)(a^{\dagger}a\rho+\rho
a^{\dagger}a-2a\rho
a^{\dagger})\nonumber\\
&-&\kappa n_T(aa^{\dagger}\rho+\rho a a^{\dagger}-2a^{\dagger}\rho
a)\nonumber\\
&-&\Gamma\sum_{i=a,b}(|e\rangle_i\langle e|\rho+\rho
|e\rangle_i\langle e|-2|g\rangle_i\langle e|\rho|e\rangle_i\langle
g|).
\end{eqnarray}
Here $\Gamma$ describes the atom decay rate and $\kappa$ stands
for the cavity leakage  rate. We do not explicitly specify the
white noise, but its intensity $n_T$ refers to its effective
particle number. To simplify the representation, now we turn to an
interaction picture with respect to
$H_0=\sum_{i=a,b}\frac{\omega}{2}\sigma_i^z+\omega a^{\dagger}a$.
After this transformation, the Liouvillean part remains unchanged,
while the Hamiltonian part is now given by
\begin{equation}
H_I=\sum_{i=a,b} g_i(|g\rangle_i\langle e|a^{\dagger}+h.c.),
\end{equation}
where that the atom-cavity coupling is on resonance was assumed.
The analytical solution to the equation (1) is extremely tedious.
To make the physical interpretation clear, we introduce two new
effective atomic modes one of which will be decoupled from the
cavity mode. The two collective atomic modes is given by the
following definition
\begin{equation}
\sigma_A^+=\frac{g_a\sigma_a^++g_b\sigma_b^+}{\sqrt{g_a^2+g_b^2}},
\sigma_B^+=\frac{g_b\sigma_a^+-g_a\sigma_b^+}{\sqrt{g_a^2+g_b^2}},
\end{equation}
with $\sigma_i^+=|e\rangle_i\langle g|$, together with
$\sigma_i^-=(\sigma_i^+)^{\dagger}$, and $\sigma_i^z$ represent
the pauli operators for the atom $i$. In terms of these new
operators, the Hamiltonian and Liouvillean part of the master
equation are given by
\begin{equation}
H_I=g(|g\rangle_A \langle e|a^{\dagger}+h.c.),
\end{equation}
where $g=\sqrt{g_a^2+g_b^2}$, and
\begin{eqnarray}
{\cal L}(\rho)&=&-\kappa(n_T+1)(a^{\dagger}a\rho+\rho
a^{\dagger}a-2a\rho
a^{\dagger})\nonumber\\
&-&\kappa n_T(aa^{\dagger}\rho+\rho a a^{\dagger}-2a^{\dagger}\rho
a)\nonumber\\
&-&\Gamma\sum_{i=A,B}(|e\rangle_i\langle e|\rho+\rho
|e\rangle_i\langle e|-2|g\rangle_i\langle e|\rho|e\rangle_i\langle
g|).
\end{eqnarray}
Note that the sum in the last line of Eq.(7) is taken over the two
NEW modes. The transformation between the resulting atom $a,b$ and
the collective modes $A, B$ is clear. For example, both the
resulting atoms $a$ and $b$ in its ground state
$|g\rangle_a|g\rangle_b$ can be equivalently expressed in terms of
$|g\rangle_A|g\rangle_B$ and $|e\rangle_a|e\rangle_b$ likewise.
The new master equation Eq.(7) shows us that we have one mode
(mode B) which is completely decoupled from the Hamiltonian
dynamics and is purely damped under the Liouvillean dynamics, this
is a consequence of the transformation from the resulting atoms to
the collective modes. The mode $B$ then will not be populated in
steady state irrespective of its initial states. In other words,
if the mode $B$ is in its ground state initially, it will remain
on that forever. Therefore, we begin our investigations with both
collective modes $A$ and $B$ in the ground state
$|g\rangle_A|g\rangle_B$. As the mode $B$ will then never be
populated, we disregard that mode in the following discussions.
Apart from the above assumption, we discuss the entanglement
generation here only for the case of no photon in the cavity
initially, this is relevant to the topics under our consideration,
i.e., study the role of the white noise in the entanglement
generation. To understand the origin of the generation of
entanglement from white noise, let us begin by considering a
special case of perfect cavity, i.e., $\kappa=0$. For the general
form of the reduced density matrix of the atom system
$$\rho_{atom}=\rho_{ee}|e\rangle\langle
e|+\rho_{eg}|e\rangle\langle g|+\rho_{ge}|g\rangle\langle
e|+\rho_{gg}|g\rangle \langle g|,$$it is easy to check that
$\rho_{eg}=\rho_{ge}=0$ in this situation, where
$\rho_{xy}(x,y=g,e)=Tr_c(|y\rangle \langle x|\rho)$ and $Tr_c$
denotes trace over the cavity mode. Noticing $|e\rangle_A\langle
e|=\frac{g_a^2|e\rangle_a\langle e|+g_b^2|e\rangle_b\langle
e|}{g_a^2+g_b^2}$, and $|g\rangle_A\langle g|$ has a similar form,
we conclude that for perfect cavity with no initial photon inside
the cavity, the atoms remain separable at all the times. Moreover
this conclusion holds even for $\kappa\neq 0$ with $n_T=0$. In
this case, the cavity mode will never be populated if there is no
photon in the cavity initially. As the cavity mode act as a data
bus, no photon exist inside the cavity means the data bus is idle.
This conclusion will be changed if we increase the value of the
thermal noise intensity $n_T$. For a limiting case of $g=0$ and
$\Gamma=0$, the stationary state of the cavity field is no longer
the vacuum, instead, it takes\cite{gardiner}
$$\rho_c=\sum_n|n\rangle\langle
n|(\frac{n_T}{1+n_T})^n\frac{1}{1+n_T}.$$ The  active data bus
indicate that the two atoms inside the cavity will be entangled
together.

We will now show in a numerical way that the envisioned idea we
have described above can be realized in the atom-cavity system by
properly chosen parameters. We will choose the Wootters
concurrence as the entanglement measure \cite{wootters},
$$ c(\rho)=max\{0,\lambda_1-\lambda_2-\lambda_3-\lambda_4\},$$
where the $\lambda_i$ are the square roots of the eigenvalues of
the non-Hermitian matrix $\rho\tilde{\rho}$ with
$\tilde{\rho}=(\sigma_y\otimes
\sigma_y)\rho^*(\sigma_y\otimes\sigma_y)$ in decreasing order. The
Wootters concurrence gives an explicit expression for the
entanglement of formation, which quantifies the resources needed
to create a given entangled state. The typical behavior of the
entanglement in the system is illustrated in figure 2.
\begin{figure}
\includegraphics*[width=0.95\columnwidth,
height=0.6\columnwidth]{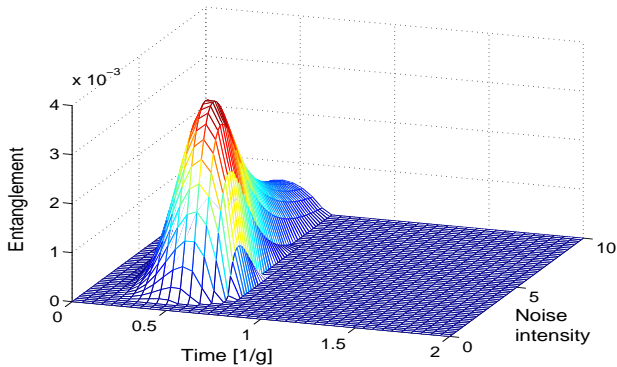} \caption{ Wootters concurence of
the two-atom system as a function of the intensity of the white
noise  $n_T$ and time $t$. The chosen parameters are
$\omega=\omega_f$, $g_a=g_b=1$, $\kappa=2$ and $\Gamma=0.2$. Note
that the entanglement arrives at its maximum for an intermediate
noise intensity. } \label{fig2}
\end{figure}
There we have plotted the amount of entanglement of the joint
state of the two atoms as a two-variable function of the intensity
of the noise $n_T$ and time $t$. The chosen parameters are
$\omega=\omega_f$, $g_a=g_b=1$, $\kappa=2$ and $\Gamma=0.2$. We
want to stress that our simulation is presented for Eq.(1), i.e.,
the original master equation for the atom-cavity system, and as we
mentioned above the initial state of the global system is
$|g\rangle_a|g\rangle_b|0\rangle_c$ in our simulation, and we cut
off the intra-cavity photon number at a value of 5. Note that for
any value of $t$ in the region of entanglement $\neq 0$, the
behavior of the amount of entanglement between the two atoms is
non-monotonic, it increase to a maximum value for an optimal
intensity of the noise and then decrease towards zero for a
sufficiently large intensity. Physically, to get non-zero amount
of entanglement, the cavity mode must be populated( or be excited
) at any value of time $t$. For the two limiting case of either
$n_T=0$ or $n_T\rightarrow \infty$, however, the data bus remains
idle for all the times. Thus the amount of entanglement equals
zero. It is also worthwhile to study the dependence of
entanglement on both the intensity of the noise and the cavity
decay rates. In figure 3, we present those for $\omega=\omega_f$,
$g_a=g_b=1$, and $\Gamma=0.2$.
\begin{figure}
\includegraphics*[width=0.95\columnwidth,
height=0.6\columnwidth]{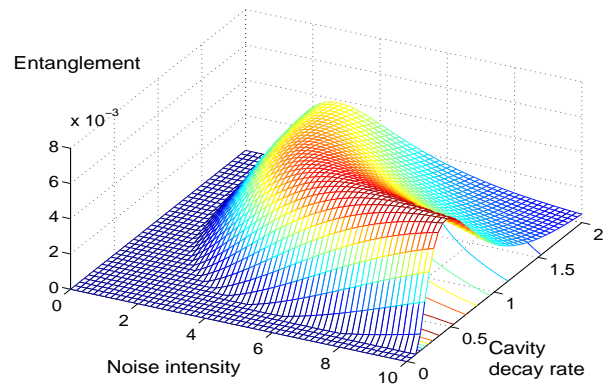} \caption{ The dependence of
amount of entanglement on the noise intensity $n_T$ and the cavity
decay rate $\kappa$ for a value of time $t=1/(2g)$. The parameters
chosen are  $\omega=\omega_f$, $g_a=g_b=1$, and $\Gamma=0.2$. The
system exhibits a resonance for both $n_T$ and $\kappa$ at an
intermediate value.} \label{fig3}
\end{figure}
It seems that the amount of entanglement  is fixed for a specific
value of $n_T\kappa$. It is interesting to note that the amount of
entanglement behave as a monotonic function of the atom decay rate
$\Gamma$. As figure 4 shows, this is quite different from the case
presented in Ref. \cite{plenio3}, where the cavity decay can
assist themselves to  entangle together, and the entanglement is
generated among themselves then.

\begin{figure}
\includegraphics*[width=0.95\columnwidth,
height=0.6\columnwidth]{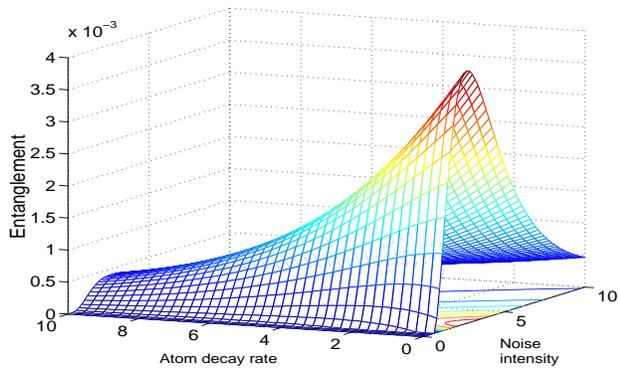} \caption{ Amount of entanglement
versus the noise intensity $n_T$ and the atomic spontaneous
emission rate $\Gamma$ for a value of time $t=1/(2g)$. We choose
the parameters $g_a=g_b=1, \omega_f=\omega$ and $\kappa=2$ for
this plot. The entanglement of the two atoms is a monotonic
function of $ \Gamma$ for any value of $n_T$, this is quite
different from the case presented in Ref.\cite{plenio3} }
\label{fig4}
\end{figure}

The noise-assisted entanglement preparation is somehow
reminiscient of the well known phenomenon of stochastic resonance
\cite{gammaitoni,huelga,buchleitner}, where the response of a
system to a periodic force can be enhanced in the presence of an
intermediate amount of noise. A related effect that cavity decay
can assist the generation of squeezing has been found recently
\cite {nha}, there they show that the squeezing effect is enhanced
as the damping rate of the cavity is increased to some extent.
However, the pumping field amplitude is required to be inversely
proportional to the damping rate for the optimal squeezing. This
is similar to our results shown in figure 3.

To sum up, we have described an experimental situation where
entanglement between two atomic systems can be prepared with
assistance of the white  noise. The entanglement measured by the
Wootters concurence is maximized for intermediate values of the
cavity decay rate and the intensity of the white noise, while it
is a monotonic function of the atomic spontaneous emission rate.
Recall that the atomic decay itself can not induce entanglement
among the atoms, even if at finite temperatures, we conclude that
the coupling between the data bus and the white noise is the
origin of the generation of the entanglement. The phenomenon of
white noise-assisted entanglement generation is not a rare
phenomenon, it resembles the phenomenon of stochastic resonance.
However, this discovery \cite{plenio3} is really valuable because
it sheds new light on the constructive role that noise may play in
quantum information processing. In contrast with the results in
Ref.\cite{plenio3}, the proposal presented here is for the
entanglement generation between two two-level atoms. For such a
two-qubit system, any amount of entanglement, even if very small,
is distillable \cite{yi}, and therefore the entangled atoms are
useful for quantum information processing. The entanglement among
many atoms also can be created by the same manner, the results
will be
presented elsewhere. \ \ \\
\ \ \\
{\bf ACKNOWLEDGEMENT:}\\ This work is supported
by EYTP of M.O.E, and NSF of China.\\

\end{document}